\documentclass[twocolumn]{jpsj3}
\usepackage{times}
\usepackage{color}

\newcommand{\crs}{CeRu$_2$Si$_2$}
\newcommand{\yrs}{YbRh$_2$Si$_2$}

\begin{document}

\title{Quantum criticality and Lifshitz transition in the Ising system CeRu$_2$Si$_2$: Comparison with YbRh$_2$Si$_2$}

\author{A.~Pourret,$^1$ D.~Aoki,$^{1,2}$ M.~Boukahil,$^1$ J.-P.~Brison,$^1$ W.~Knafo,$^3$ G.~Knebel,$^1$ S.~Raymond,$^1$  M.~Taupin,$^1$ Y.~\={O}nuki,$^4$ and J.~Flouquet$^1$} 

\inst{$^1$SPSMS, UMR-E CEA / UJF-Grenoble 1, INAC, Grenoble, F-38054, France\\
$^2$IMR, Tohoku University, Oarai, Ibaraki 311-1313, Japan\\
$^3$Laboratoire National des Champs Magn´etiques Intenses, UPR 3228,
CNRS-UJF-UPS-INSA, 143 Avenue de Rangueil, 31400 Toulouse, France\\
$^4$Faculty of Science, University of the Ryukyus, Nishihara, Okinawa 903-0213, Japan}


\abst{New thermoelectric power (TEP) measurements on prototype heavy-fermion compounds close to magnetic quantum criticality are presented. The highly sensitive technique of TEP is an unique tool to reveal Fermi surface instabilities, referred here as Lifshitz transitions. The first focus is on the Ising CeRu$_2$Si$_2$ series. Doping \crs\ with Rh produces a decoupling between the first order metamagnetic transition and the pseudo-metamagnetism observed in the pure compound. Comparison is made with the case of YbRh$_2$Si$_2$ which is often considered as the archetype of local quantum criticality by contrast to CeRu$_2$Si$_2$, taken as an example of spin-density wave criticality. 
Up to now for ferromagnetic materials showing ferromagnetic wings, no simple case appears where the Fermi surface is preserved between the ferromagnetic and paramagnetic phases. An open issue is the consequence of Lifshitz transitions on superconductivity in these multiband systems.}

\kword{\crs , \yrs , metamagnetism, pseudo-metamagnetism, Lifshitz transition, thermoelectric power, neutron scattering}

\maketitle

\section{Introduction}
The open question in the study of magnetic quantum criticality, either ferromagnetic (FM) or antiferromagnetic (AF), in heavy-fermion compounds is its feedback on the Fermi surface (FS). Due to the renormalized low characteristic energy, the determination of the full FS remains a ''tour de force" as it requires often very low temperatures, a simultaneous scan of the magnetic (FM or AF) as well as the paramagnetic (PM) phases. Focusing on highly anisotropic magnetic crystals of Ising-type is an excellent starting point as magnetic field scans along the hard axis gives the possibility to remain in the same phase, despite reaching high magnetic field strength necessary detect the Fermi surface (FS) orbits. This opens the possibility to unambiguous quantum oscillation experiments. To select a given ground state, additional tools will be given by pressure or doping. 

The limitation of quantum oscillations methods as the de Haas van Alphen (dHvA) technique is that they require a finite magnetic field window which precludes to realize a continuous derivation of the FS evolution. An excellent complementary probe is given by the  thermoelectric power \cite{Varlamov1989}. Its strength to detect FS singularities has been demonstrated in the study of Lifshitz singularities three decades ago \cite{Lifshitz1960}. Here we will classify FS reconstructions by the label of Lifshitz transitions despite the fact that in a complex multiband FS, the motion of one band will react on the others.

\section{The Antiferromagnetic -- Paramagnetic Quantum Phase Transition in the Ising familly CeRu$_2$Si$_2$}

\begin{figure}
\begin{center}
\includegraphics[width=7cm]{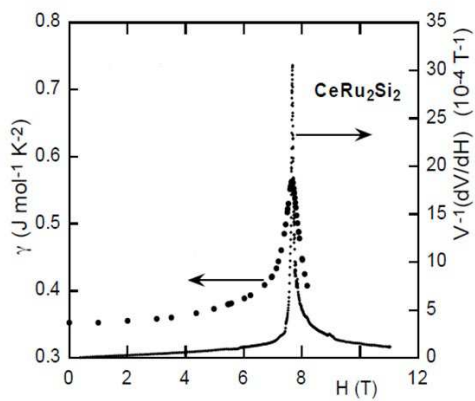}
\end{center}
\caption{Pseudometamagnetism of \crs\: Field dependence of the Sommerfeld coefficient $\gamma (H)$. A huge effect also appears in the magnetostriction $\frac{1}{V}\frac{dV}{dH}$ \cite{Flouquet2005,Flouquet2002,Aoki2011c}.}
\label{f1}
\end{figure}

Let us focus  on the tetragonal \crs\ series, where the Ce ions have strong Ising character. The anisotropy ratio between the initial susceptibility along the easy ($c$ axis) and the hard axis (basal plane) reaches 20 at low temperature \cite{Haen1992}. The pure lattice \crs\ is already in a PM ground state. The effective critical pressure $p_c$ characteristic of the switch from AF to PM is estimated at a slightly negative effective pressure of a few kbar \cite{Flouquet2005, Flouquet2002}. 

One of the spectacular phenomena is that, under magnetic field $H_m \approx 7.8$~T applied along the easy axis  a sharp continuous pseudo-metamagnetic transition suddenly occurs for a critical value of the magnetization $M=M_{cr} \approx 0.5$~$\mu_B$/Ce \cite{Flouquet2005, Flouquet2002, Aoki2011c}. At $H_m$ a sharp enhancement is observed in the Sommerfeld coefficient of the specific heat $\gamma (H)$ and an even sharper is observed in the magnetic susceptibility $\chi (H)$ due to the additional contribution of the magnetostriction (see Fig.~\ref{f1}).

\begin{figure}
\begin{center}
\includegraphics[width=7cm]{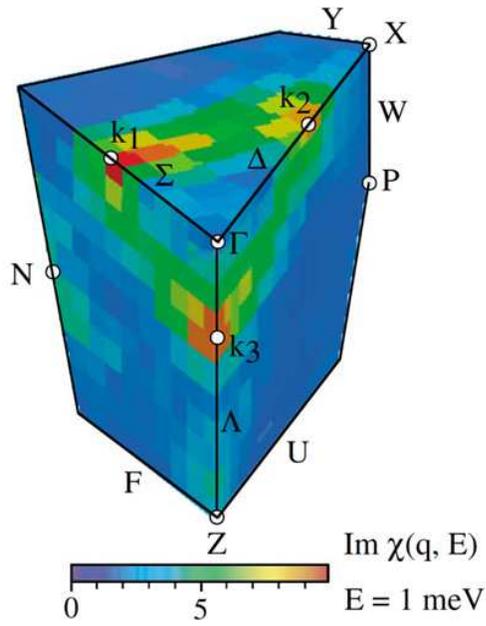}
\end{center}
\caption{(Color online) Appearance of three AF hot spots in the reciprocal space of \crs\ from inelastic neutron scattering data of the dynamical susceptibility \cite{Kadowaki2004}.}
\label{f2}
\end{figure}

\begin{figure}
\begin{center}
\includegraphics[width=7cm]{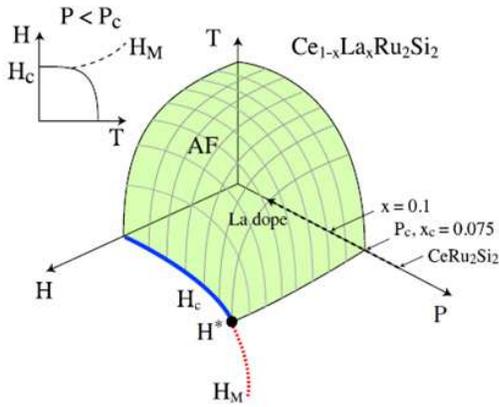}
\end{center}
\caption{(Color online) $(T,H,p)$ phase diagram of \crs\ series derived from pressure $p$ or doping studies with La and Ge.}
\label{f3}
\end{figure}

Inelastic neutron scattering experiments at $H=0$ on \crs\ point out  that three AF hot spots exist at $\mathbf{k_1} = (0.31,0,0)$, $\mathbf{k_2} = (0.31,31,0)$, $\mathbf{k_3} = (0,0,0.35)$ (see Fig.~\ref{f2}, taken from Ref.~\citen{Kadowaki2004}). Doping with La or Ge leads to AF order at the wave vector $\mathbf{k_1}$ (see Fig.~\ref{f3}) \cite{Quezel1988, Mignot1991}, while doping with Rh induces AF order at $\mathbf{k_3}$ \cite{Kawarazaki1997}. In pure \crs\ crossing through $H_m$ is accompanied by a collapse of the AF correlations, by an enhancement of the low energy FM fluctuations at $H_m$, and finally by a wavevector $k$ independent response in the polarized paramagnetic (PPM) domain \cite{Raymond,Sato2004}. As the AF inelastic signal does not show any shift at low energy as $H$ approaches $H_m$ \cite{Raymond}, it is clear that the dominant process cannot be due to a drastic change from the proximity to the AF singularity. 

\begin{figure}
\begin{center}
\includegraphics[width=7cm]{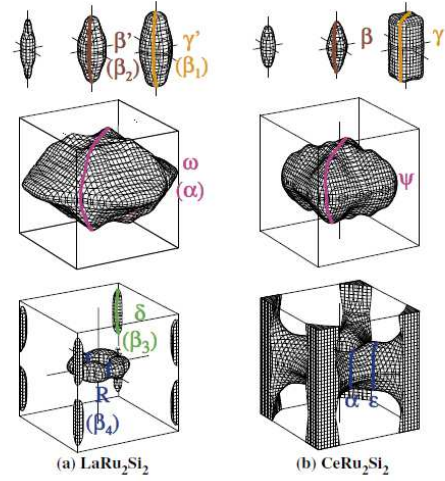}
\end{center}
\caption{(Color online) Fermi surface predicted for LaRu$_2$Si$_2$ and \crs\ assuming that the 4$f$ electrons are itinerant \cite{Flouquet2005, Aoki1992, Takashita1996}. }
\label{f4}
\end{figure}

Spectacularly, a drastic change of the FS occurs at $H_m$. At low field $H<H_m$ \cite{Aoki1992, Takashita1996}, the Fermi surface of \crs\ \cite{Aoki1992, Takashita1996, Lonzarich1988} has been almost fully determined experimentally and even a quantitative agreement is found in band structure calculations with an itinerant treatment of the 4$f$ electrons \cite{Suzuki2010}. Objections of a localized picture above $H_m$ have been made by different authors (see in ref.~\citen{Flouquet2005}). The link with a Lifshitz transition, where one of the spin-split FS sheets vanishes, was first pointed out in ref.~\citen{Daou2006}. Above $H_m$, only few parts of the FS have been determined pushing to the image that in the PPM state the measured orbits coincides with those detected for LaRu$_2$Si$_2$ \cite{Aoki1992, Takashita1996}. From the quantum oscillation evolution through $H_m$ the average effective mass of the electron orbits never exceeds that of the hole orbits. Figure~\ref{f4} represents the FS obtained in band structure calculation for LaRu$_2$Si$_2$ and \crs\ for $H<H_m$.

It is interesting to notice that in the AF phase close to the critical doping $x>x_c$ where the switch from AF to PM in the alloys CeRu$_2$(Si$_{1-x}$Ge$_x$)$_2$ appears, the FS is similar to that detected in the PM regime of \crs . Thus, the FS instability from localized to itinerant character is here decoupled from magnetic quantum criticality; it occurs deep inside the AF ordered phase at a concentration $x_a > x_c$,  in excellent agreement with the recent theoretical work of Ref.~\citen{Okane2009}. Furthermore, from ARPES spectroscopy experiments, it has been even stressed that this 4$f$ itineracy occurs for $x < x_a$ even in the PM regime above the N\'eel temperature \cite{Hoshino2013}.

\begin{figure}
\begin{center}
\includegraphics[width=7cm]{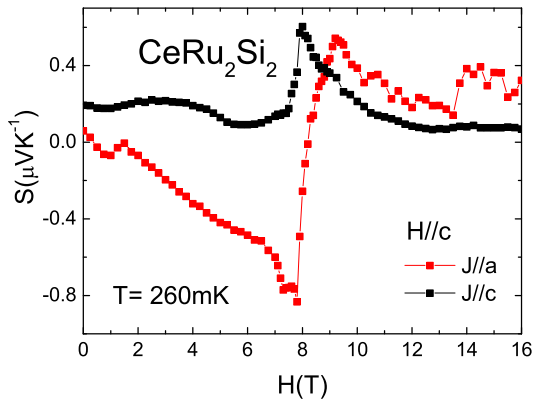}
\end{center}
\caption{(Color online) Field dependence of the TEP $(S)$ at 260 mK for the transverse and longitudinal configurations of \crs . }
\label{f5}
\end{figure}

Doping  with La or Ge is dominated mainly by the expansion of the volume of the unit cell leading to an AF order stabilized at a transverse wave vector. Pressure experiments on La doped crystals have clearly shown that the pseudo-metamagnetic transition of \crs\ at $H_m$ is reminiscent of the first order metamagnetism which occurs at $H_c$ in the AF domain, and that furthermore $H_c$ terminates at a field quantum critical end point $H_{QCEP}$ \cite{Flouquet2005,Weickert2010}. If a field sweep is realized just near $x_c$ close to $T_N$,  two characteristic fields  $H_c$ and $H_m$ are revealed. A key point is that metamagnetism and pseudometamagnetism occur at a given value of the magnetization $M=M_{cr}$. This is a direct illustration that the field induced transitions  from AF to PPM at $H_c$ or from PM to PPM at $H_m$ are driven by a FS instability. 

\begin{figure}
\begin{center}
\includegraphics[width=7cm]{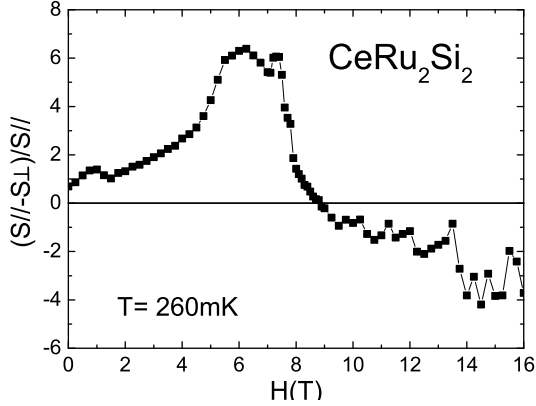}
\end{center}
\caption{(Color online) Field dependence of the anisotropy of the TEP $\frac{S_\perp -S_\parallel}{S_\parallel}$ of \crs\ at 260 mK \cite{Boukahil2013}. 
}
\label{f6}
\end{figure}

Two decades ago, previous TEP ($S$) measurements on \crs\ \cite{Amato1989}  have shown that a  drastic change of $S(H)$ occurs at $H_m$. Improved crystal qualities pushed to a new generation of experiments \cite{Pfau2012, Boukahil2013}. Here we will focus on the TEP through the pseudo-metamagnetic transition with a transverse thermal gradient applied in the basal plane ($\nabla T \parallel a$) and along the $c$ axis ($\nabla T \parallel c$) \cite{Boukahil2013}. Figure \ref{f5} shows the field dependence of $S$ at $T = 260$~mK through $H_m$ for the different configurations ($\nabla T \parallel a$) and ($\nabla T \parallel c$). The striking point is the change of sign in the TEP anomaly between the two configurations at $H_m$  (see Fig.~\ref{f6}). This anisotropic $H$ response may suggest differences in the hybridization. It can also be a consequence of the different scatterings for the longitudinal and transverse response, as observed in the magnetoresistivity, since the total TEP is the sum of the TEP of each band weighted by their respective electrical conductivity. One interest of this work was to succeed following the TEP on a well documented case, where a FS instability and its reconstruction is established. Furthermore, by improving the crystal quality, it was possible to detect quantum oscillations of the TEP too. The observed frequencies are in excellent agreement with the different low frequencies detected by dHvA experiments on both sides of $H_m$ \cite{Aoki1992,Takashita1996}. 

\section{Decoupling of $H_c$ and $H_m$ in \crs\ via Rh doping}

\begin{figure}
\begin{center}
\includegraphics[width=7cm]{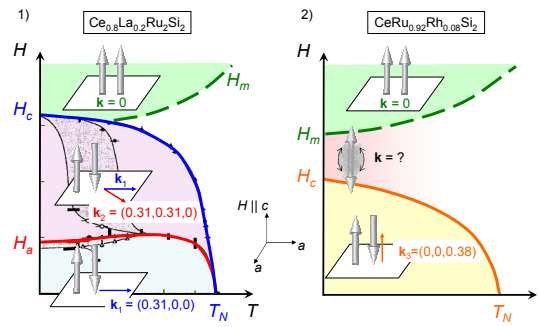}
\end{center}
\caption{(Color online) Comparison of the $(H,T)$ phase diagram for La and Rh doped AF phase \cite{Aoki2012}.}
\label{f7}
\end{figure}

\begin{figure}
\begin{center}
\includegraphics[width=7cm]{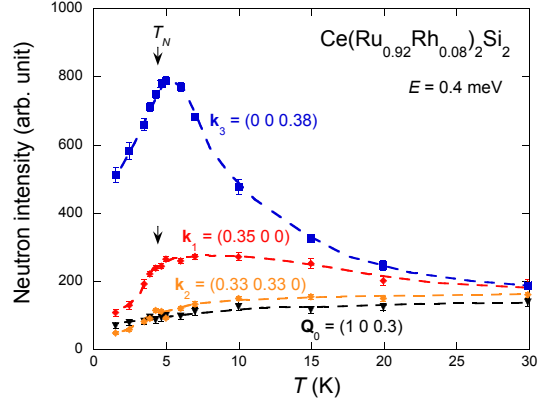}
\end{center}
\caption{(Color online) Inelastic neutron scattering response of Ce(Ru$_{0.92}$Rh$_{0.08}$)$_2$Si$_2$ as function of temperature at different wave vectors for an energy transfer $E=0.4$~meV \cite{Knafo2013}.}
\label{f8}
\end{figure}

Doping \crs\ with Rh leads due to a mismatch of the \crs\ and CeRh$_2$Si$_2$ lattice parameters \cite{Aoki2012} which is strong enough to modify the FS so that the dominant hot spot becomes the longitudinal wavevector $\mathbf{k_3} \approx (0,0,0.38)$ \cite{Aoki2012} as shown in Fig.~\ref{f7}. Figure \ref{f8} shows the inelastic neutron scattering response measured on the triple axis spectrometer IN12 at the ILL at an energy of 0.4~meV for an Rh doping of 8\%. The strong enhancement at $T_N$ of the magnetic fluctuations at the wave vector $\mathbf{k_3}$ indicates that these order parameter fluctuations are critical and drive the AF transition. Similarly, critical order parameter fluctuations at the wave vector $\mathbf{k_3}$ have also been observed at the N\'eel temperature of La doped \crs\ \cite{Knafo2009}. 
Applying a magnetic field along the $c$ axis leads to the disappearance of long range magnetic order at the magnetic field $H_c \approx 2.5$~T with no further evidence of long range order above $H_c$. However, a pseudo-metamagnetic transition occurs at $H_m \approx 5.5$~T $>H_c$ and is reminiscent of the phenomena just described above on La or Ge doped \crs\  (Fig.~\ref{f7}). 

\begin{figure}
\begin{center}
\includegraphics[width=7cm]{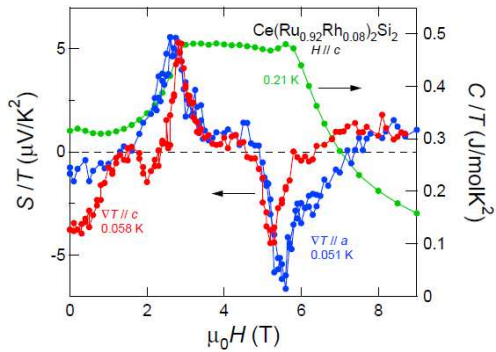}
\end{center}
\caption{(Color online) Field dependence of Ce(Ru$_{0.92}$Rh$_{0.08}$)$_2$Si$_2$ of the TEP at $T \sim 55$~mK and $C/T$ at 0.21~K \cite{Machida2013}.}
\label{f9}
\end{figure}

TEP experiments where realised on the 8\% Rh doped compound in the transverse ($\nabla T \parallel a$) and longitudinal ($\nabla T \parallel c$) configuration \cite{Machida2013}. Figure \ref{f9} shows the field response $S(H)$ at $T = 55$~mK. Two singularities appear on crossing $H_c$ and $H_m$, but their respective amplitudes depend weakly  of the chosen configuration. This behavior is quite different to that observed for the pure \crs\ case, as doping by Rh leads to a rather high residual resistivity. Thus the quasiparticles are far to feel their quantum FS. In absence here of quantum oscillation experiments, these TEP data suggest two different FS reconstructions at $H_c$ and $H_m$. It is interesting to note that in this Rh doped compound again, $M$ reaches a value near $M_{cr}\approx 0.5$~$\mu_B$/Ce at $H_m$ as in the pure \crs . 

\section{\yrs : Another Case of Zeeman Decoupling without Metamagnetism}

\begin{figure}
\begin{center}
\includegraphics[width=7cm]{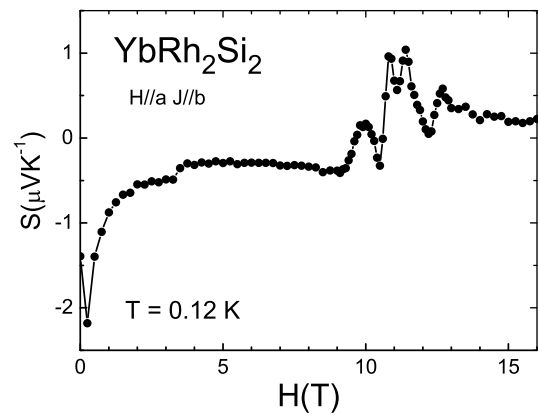}
\end{center}
\caption{(Color online) Evidence of Lifshitz transitions in \yrs\ at $H_0 \sim 9.5$~T far above the AF critical field $H_c \sim 0.06$~T \cite{Pourret2013}.}
\label{f10}
\end{figure}

In this conference, extensive discussions have been given on the situation of \yrs\ ($T_N = 70$~mK) on both sides of the critical field $H_c$ where the AF order is suppressed. \yrs\ has also a large magnetic anisotropy but with its easy axis along the basal plane and its hard axis along the $c$ axis \cite{Gegenwart2002}.  Thus, the  situation is very different from the previous Ising-type example. Anyway, increasing the magnetic field along the easy plane also leads to reach a critical value of the magnetization ($M \approx 1 \mu_B$)/Yb above which the FS becomes unstable. This phenomenon has been first pointed out in dHvA experiments \cite{Rourke2008}. The FS instability is  clearly seen in recent TEP measurements \cite{Pfau2013,Pourret2013}, as shown in Fig.~\ref{f10}. Furthermore a cascade of FS reconstructions emerges in good agreement with the multiband structure of the complex FS \cite{Pfau2013}. By contrast to \crs\, the phenomenon is not linked to metamagnetism and an enhancement of $\gamma (H)$. In difference to \crs , in \yrs\ $\gamma (H)$ has a monotonous $H$ decrease on crossing the field cascade at $H \sim 10$~T. The common point with the previous \crs\ case is the FS reconstruction driven by a critical value of the magnetic polarization. Similar phenomena had been invoked for the polarized phase of URu$_2$Si$_2$ \cite{Pourret2013a} and of UCoGe \cite{Malone2012}. In \yrs\ it was even proposed that a FS reconstruction at $H_c$ may not be due to local criticality but to a Zeeman driven Lifshitz transition \cite{Hackl2011}.  

The main trend is to classify the \crs\ series as prototype of spin-density wave instability \cite{Knafo2009} which was already introduced three decades ago, and \yrs\ as a fresh example of the novel approach of local quantum criticality \cite{Loehneysen2007, Gegenwart2008}. It may be worthwhile to note that Yb heavy fermion compouns (HFC) are not straightforward the hole analog of the Ce electron Kondo lattice. Differences in spin-orbit coupling and in hybridization \cite{Flouquet2012} make that the band structures of the Yb HFC differs strongly from that of the Ce HFC as pointed out by comparing CeRh$_2$Si$_2$ to \yrs\ \cite{Knebel2006a}. The consequence of an increasing local character of the Yb ion is illustrated by the weaker dispersion of the effective masses for Yb than for Ce HFC between heavy and light carriers of their multiband Fermi sheets. Thus, it is not  obvious that there is a clear separation between Ce and Yb HFC criticality. It is worthwhile to emphasize that in \crs\ already at 60~K which is far above the postulated Kondo temperature of 25~K, antiferromagnetic correlations are visible \cite{Flouquet2005} in good agreement with the theoretical statement that Kondo screening disappears close to $p_c$ but the coupling to spin fluctuations remains \cite{Maebashi2005}. To reach a consensus, new Yb material which could open the possibility of a full determination of of the FS must emerge. Let us notice that TEP of \yrs\ through $H_c$ may not agree with a local quantum critical scenario \cite{Machida2012}. 

\begin{figure}
\begin{center}
\includegraphics[width=7cm]{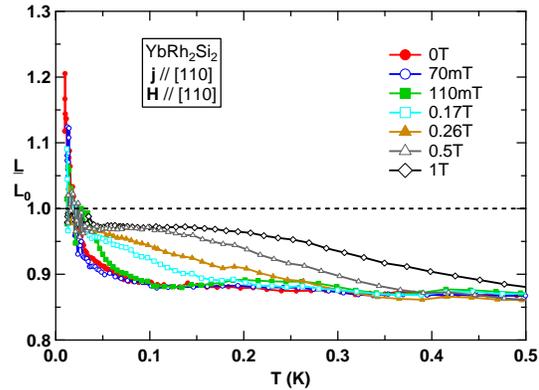}
\end{center}
\caption{(Color online) Lorentz number derived from transport measurements down to 8~mK for $H \parallel [110]$ ($H_c = 66$~mT). To extract $L$ down to $T\to 0$ is still an open problem \cite{Taupin2013}.}
\label{f11}
\end{figure}

Recently, four different groups (Dresden \cite{Pfau2012a}, Tokyo \cite{Machida2013a}, Sherbrooke \cite{Reid2013}, Grenoble \cite{Taupin2013}) have realized thermal conductivity experiments on \yrs\ as a possible test of the  breakdown of a ''conventional" Fermi liquid description via the violation of the Wiedemann-Franz law at the critical field $H_c$. A claim of this violation has been made in ref.~\citen{Pfau2012a} whereas the persistence of the Wiedemann-Franz law has been stessed in refs.~\citen{Machida2013a,Reid2013}. A definite conclusion require now to achieve the very low temperature limit.  The difficulty to extract the quasiparticle Lorentz number $L$ at $T \to 0$ is linked to the very low value of the N\'eel temperature ($T_N = 70$~mK) correlated with the emergence of an additional extra contribution \cite{Pfau2012a} (caused by magnons or exotic coherent excitations) which will only collapse far below 8~mK, the lowest temperature achieved in the Grenoble experiment\cite{Taupin2013} (see Fig.~\ref{f11}). The final experimental test may become extremely hard as on cooling an additional weak phase transition occurs at $T=14$~mK and a new magnetically ordered phase below 2.2~mK \cite{Schuberth2009, Schuberth2013}. 

Here we point out that unfortunately the reports on HFC criticality have been mainly focused on the PM side of the instability. There is a quasi-complete omission of quantitative considerations on the AF side, notably on the specific heat anomalies at $T_N$ when $T_N \to 0$ and on the characteristic excitations  deep below $T_N$. Even if the FS may coincide with the itinerant picture at $T \to 0$, the duality between itinerant and local characters of the 4$f$ electrons may lead to a fascinating interplay on cooling with signatures on both sides of  $T_N$ and $H_c$. The new intriguing feature in \yrs\ is that the extra contribution survives for fields far above $H_c$ which collapses only at high magnetic fields $H \gg H_c$ \cite{Taupin2013} (see Fig.~\ref{f11}).  
Thus the experimental data may be interpreted without any particular anomaly at the critical field $H_c$ and without violation of the Wiedemann-Franz ratio.

\section{Ferromagnetic Wings and Fermi Surface Instability}

Ferromagnetic wings have been clearly detected in the uranium HFC UGe$_2$ \cite{Taufour2010} and UCoAl \cite{Aoki2011b}.  At least for UGe$_2$, it is clear that the switch from the PM to FM phases is associated to a FS reconstruction \cite{Aoki2011b, Kotegawa2011}. For the second case, UCoAl, the situation is still not clear as  the FS has not been detected up to now.  At least there are indications from preliminary band structure calculation that a FS reconstruction may occur also in UCoAl at the border of the PM and FM phases\cite{Harima2013}. Recent TEP measurements show a large anomaly at the transition from PM to FM under field \cite{Palacio-Morales2013}. Comparison between experiment and theoretical calculations of the FS is a key point. It is  up to now not proved that the band-structure approach gives a sound description of the FM FS when the FM sublattice magnetization decreases on approaching the FM--PM switch. Here it is intriguing that for the FM superconductor UCoGe, despite a FM moment of only 0.05 $\mu_B$ a drastic FS reconstruction is predicted between the FM and PM phases \cite{Samsel-Czekala2010}. 

An additional interest is to investigate the consequences of a Lifshitz transition on superconducting pairing as a Lifshitz transition induces a drastic change of the density of states. In a single band image this problem was considered in order to explain the field reentrance of the superconductivity  at $H = H_R$ in the FM superconductor URhGe \cite{Yelland2011}, when the field is applied along the hard $b$ axis. The disappearance of one Shubnikov de Haas frequency on approaching $H_R$ leads to a drop of the corresponding quasiparticle velocity and is thus a key source for the enhancement of the upper superconducting critical field \cite{Yelland2011}.  However, as the  full  FS of this multiband system will react, a quantitative comparison requires informations on the field variation of other FS orbits. To our knowledge, there is a lack of theoretical descriptions of Lifshitz transitions and superconductivity. 

\section{Conclusions}

To summarize, a new generation of TEP measurements on high quality single crystals of \crs\ series demonstrate the power of the method to detect Lifshitz transitions in heavy fermion metals. Simultaneous information on the average TEP power response and on the additional quantum oscillation response directly linked to the quasiparticle quantum orbit of the FS can be achieved. Comparison is made with the case of Yb HFC by focusing on \yrs . A common point between \crs\ and \yrs\ is the $H$ crossing through  a critical value of the magnetic polarization, which forces a Lifshitz-like transition of the FS. Emphasis is made on the necessity to pay attention on the coherent state achieved below the magnetic ordering temperature near its collapse at $T \to 0$.

Up to now, no clear  case seems to occur where the FM wing may be separated from FS instability. Finally, a remark is made on the necessity for theoretical modeling of the interplay between FS instabilities and superconductivity.

\section*{Acknowledgements}
We thank Y.~Machida and K.~Izawa for the collaborations on the Rh doped \crs , and K.~Miyake and H.~Harima for many fruitful discussions.  This work has been supported by the French ANR project PRINCESS and the ERC starting grant NewHeavyFermion.




\end{document}